\documentclass[3p,times]{elsarticle}
\usepackage{ecrc}
\usepackage{amssymb,amsmath}
\usepackage[usenames]{color}
\usepackage{algorithm}
\usepackage{algpseudocode}
\usepackage{listings}

\volume{00}
\firstpage{1}
\journalname{Computer Physics Communications}
\runauth{Simon Catterall and Anosh Joseph}

\jnltitlelogo{\small Computer Physics Communications}

\usepackage{amssymb}
\usepackage[figuresright]{rotating}

\DeclareMathOperator{\Tr}{Tr}

\newcommand{\cA}{{\cal A}}
\newcommand{\cAb}{{\overline{\cal A}}}
\newcommand{\cF}{{\cal F}}
\newcommand{\cFb}{{\overline{\cal F}}}
\newcommand{\cD}{{\cal D}}
\newcommand{\cDb}{{\overline{\cal D}}}
\newcommand{\cQ}{{\cal Q}}
\newcommand{\cU}{{\cal U}}
\newcommand{\cN}{{\cal N}}
\newcommand{\cUb}{{\overline{\cal U}}} 
\newcommand{\KD}{{K\"{a}hler-Dirac }}
\newcommand{\bx}{{\bf x}}

\newcommand{\ba}{{\boldsymbol a}}
\newcommand{\hatbe}{\widehat{\boldsymbol e}}

\newcommand{\bR}{{\boldsymbol R}}
\newcommand{\hatbmu}{\widehat{\boldsymbol {\mu}}}
\newcommand{\vx}{ {\bf x} }
\newcommand{\vn}{ {\bf n} }

\def\nn{\nonumber}
\def\bec{\begin{center}}
\def\eec{\end{center}}
\def\beq{\begin{equation}}
\def\eeq{\end{equation}}
\def\bea{\begin{eqnarray}}
\def\eea{\end{eqnarray}}

\begin{document}

\begin{frontmatter}

\dochead{}

\title{An object oriented code for simulating supersymmetric Yang--Mills theories}
\author{Simon Catterall}\ead{smcatterall@gmail.com}
\address{Department of Physics, Syracuse University, Syracuse, NY 13244, USA}
\author{Anosh Joseph}\ead{anosh@lanl.gov}
\address{Department of Physics, Syracuse University, Syracuse, NY 13244, USA\\
Theoretical Division, Los Alamos National Laboratory, Los Alamos, NM 87545, USA\footnote{$^\dagger$ Present address.}}

\begin{abstract}
We present SUSY\_LATTICE - a C++ program that can be used to simulate certain classes of supersymmetric Yang--Mills (SYM) theories, including the well known ${\cal N} = 4$ SYM in four dimensions, on a flat Euclidean space-time lattice. Discretization of SYM theories is an old problem in lattice field theory. It has resisted solution until recently when new ideas drawn from orbifold constructions and topological field theories have been brought to bear on the question. The result has been the creation of a new class of lattice gauge theories in which the lattice action is invariant under one or more supersymmetries. The resultant theories are local, free of doublers and also possess exact gauge-invariance. In principle they form the basis for a truly non-perturbative definition of the continuum SYM theories. In the continuum limit they reproduce versions of the SYM theories formulated in terms of {\it twisted} fields, which on a flat space-time is just a change of the field variables. In this paper, we briefly review these ideas and then go on to provide the details of the C++ code. We sketch the design of the code, with particular emphasis being placed on SYM theories with ${\cal N} = (2, 2)$ in two dimensions and ${\cal N} = 4$ in three and four dimensions, making one-to-one comparisons between the essential components of the SYM theories and their corresponding counterparts appearing in the simulation code. The code may be used to compute several quantities associated with the SYM theories such as the Polyakov loop, mean energy, and the width of the scalar eigenvalue distributions.
\end{abstract}

\end{frontmatter}

\section{Introduction} 
\label{sec:intro}

The problem of formulating supersymmetric theories on lattices has a long history going back to the earliest days of lattice gauge theory. However, after initial efforts failed to produce useful supersymmetric lattice actions the topic languished for many years. Indeed a folklore developed that supersymmetry and the lattice were mutually incompatible. However, recently, the problem has been re-examined using new tools and ideas such as topological twisting \cite{Sugino:2003yb, Sugino:2004qd, Sugino:2004uv, Catterall:2004np, Catterall:2005fd, D'Adda:2005zk, Catterall:2005eh, Sugino:2006uf, Catterall:2006jw, Catterall:2006is, D'Adda:2007ax, Catterall:2007kn, Catterall:2007hk, Catterall:2008dv, Catterall:2009it}, orbifold projection and deconstruction \cite{Kaplan:2002wv, Nishimura:2003tf, Cohen:2003xe, Cohen:2003qw, Kaplan:2005ta, Unsal:2006qp, Damgaard:2007be, Damgaard:2007xi, Matsuura:2007ec, Damgaard:2008pa}, and a class of lattice models have been constructed which maintain one or more supersymmetries exactly at non-zero lattice spacing\footnote{There exist other attempts to study various supersymmetric models on the lattice. See \cite{Hanada:2007ti, Anagnostopoulos:2007fw, Azeyanagi:2008bk, Hanada:2009kz, D'Adda:2009kj, Hanada:2009hq, Hanada:2010kt, Hanada:2010gs}.}.

The availability of a supersymmetric lattice construction is clearly very exciting. For example, having a lattice construction of the well known $\cN=4$ SYM in four-dimensions is very advantageous from the point of view of exploring the connection between gauge theories and string/gravitational theories. But even without this connection to string theory, it is clearly of great importance to be able to give a non-perturbative formulation of a supersymmetric theory via a lattice path integral, in the same way that one can formally define QCD as a limit of lattice QCD as the lattice spacing goes to zero and the box size to infinity. From a practical point of view, one can also hope that some of the technology of lattice field theory such as strong coupling expansions and Monte Carlo simulation can be brought to bear on such supersymmetric theories.

In this paper, we will briefly describe the key ingredients that go into the lattice constructions of a variety of SYM theories and then provide the details of C++ code that can be used to simulate these theories. In Section \ref{sec:method-twist-SYM} we provide the general method of twisting the supersymmetries of certain classes of SYM theories to provide twisted SYM theories that are compatible with discretization on the lattice. We start with the twisted $\cN=2$ SYM in two dimensions as a warm up example and after writing down the discretization of this theory we go on to describe the twisted versions of $\cN=4$ SYM in three dimensions and $\cN=4$ SYM in four dimensions. In Section \ref{sec:simulating-sym-alg} we describe the algorithms used to simulate these resultant lattice theories: rational hybrid Monte Carlo (RHMC) algorithm to compute the fermion determinant with fractional power, leapfrog algorithm to evolve the system of equations and then a Metropolis test to accept or reject the configurations. In Section \ref{sec:overall-structure} we provide the overall structure of the C++ code and describe how the code advances by generating new configurations using RHMC algorithm, saves the field configurations after some number of Monte Carlo sweeps and measures the observables in the theory. We provide some simulation results in Section \ref{sec:sim-results}, specific to the $\cN=2$ SYM in two-dimensions. We compute the eigenvalues of the scalars of the theory, study the Pfaffian phases and the presence of fermionic sign problem in the theory and also investigate the restoration of supersymmetry by checking if the scalar supersymmetry has indeed been implemented correctly in our C++ code. We provide some conclusions and outlook in Section \ref{sec:conclu}. We also provide three appendices: \ref{sec:install} details the installation of the program, \ref{sec:file-list} lists the files included in SUSY\_LATTICE library with a brief description of their purpose and in \ref{sec:input-param} we provide a sample file with input parameters. 

We hope that this work will motivate elementary particle physicists as well as high energy computational physicists to pursue numerical studies of supersymmetric lattice theories in particular, the $\cN=4$ Yang--Mills in four dimensions.

\section{The method of topological twisting in SYM theories}
\label{sec:method-twist-SYM}

First, let us explain why discretization of supersymmetric theories resisted solution for so long. The central problem is that naive discretizations of continuum supersymmetric theories break supersymmetry completely and radiative effects lead to a profusion of relevant supersymmetry breaking counterterms in the renormalized lattice action. The coefficients to these counterterms must then be carefully fine tuned as the lattice spacing is sent to zero in order to arrive at a supersymmetric theory in the continuum limit. In most cases this is both unnatural and practically impossible -- particularly if the theory contains scalar fields. Of course, one might have expected problems -- the supersymmetry algebra is an extension of the Poincar\'e algebra, which is explicitly broken on the lattice. Specifically, there are no infinitesimal translation generators on a discrete space-time so that the algebra $\{Q,\overline{Q}\}=\gamma_a p_a$, where $a$ is the space-time index, is already broken at the classical level. Equivalently, it is a straightforward exercise to show that a naive supersymmetry variation of a naively discretized supersymmetric theory fails to yield zero as a consequence of the failure of the Leibniz rule when applied to lattice difference operators. In the last five years or so this problem has been revisited using new theoretical tools and ideas and a set of lattice models have been constructed which retain exactly some of the continuum supersymmetry at non-zero lattice spacing. The basic idea is to maintain a particular sub-algebra of the full supersymmetry algebra in the lattice theory. The hope is that this exact symmetry will constrain the effective lattice action and protect the theory from dangerous supersymmetry violating counterterms.

Two approaches have been pursued to produce such supersymmetric actions: one based on ideas drawn from the field of topological field theory \cite{Sugino:2003yb, Catterall:2004np, Catterall:2005fd} and another pioneered by David B. Kaplan. Mithat \"Unsal and collaborators using ideas of orbifolding and deconstruction \cite{Cohen:2003xe, Cohen:2003qw, Kaplan:2005ta}. Remarkably, these two seemingly independent approaches lead to the same lattice theories -- see \cite{Catterall:2007kn, Unsal:2006qp, Damgaard:2007be, Damgaard:2007eh} and the recent reviews \cite{Catterall:2009it, Giedt:2009yd, Joseph:2011xy}. This convergence of two seemingly completely different approaches to the problem leads one to suspect that the final lattice theories may represent essentially unique solutions to the simultaneous requirements of locality, gauge-invariance and at least one exact supersymmetry. In this paper, we will use the language of topological twisting to discuss these supersymmetric lattice constructions, but the reader should remember that the orbifold methods lead to the same lattice theories.

\subsection{Twisting the supersymmetries in $d$ dimensions}

The basic idea of twisting goes back to Witten in his seminal paper on topological field theory \cite{Witten:1988ze} but actually had been anticipated in earlier work on staggered fermions \cite{Elitzur:1982vh}. In our context the idea is decompose the fields of the theory in terms of representations not of the original (Euclidean) rotational symmetry $SO_{\rm rot}(d)$ but a twisted rotational symmetry $SO(d)^\prime$, which is the diagonal subgroup of this symmetry and an $SO_{\rm R}(d)$ subgroup of the R-symmetry of the theory,
\beq
SO(d)^\prime={\rm diag}(SO_{\rm Lorentz}(d)\times SO_{\rm R}(d))~.
\eeq
To be explicit, consider the case where the total number of supersymmetries is $Q=2^d$. In this case we can treat the supercharges of the twisted theory as a $2^{d/2}\times 2^{d/2}$ matrix $q$. This matrix can be expanded on products of gamma matrices:
\beq
q = \cQ I + \cQ_a \gamma_a + \cQ_{ab}\gamma_a\gamma_b + \ldots
\eeq
The $2^d$ antisymmetric tensor components that arise in this basis are the twisted supercharges and satisfy a corresponding supersymmetry algebra following from the original algebra
\bea
\cQ^2 &=& 0~, \\
\{\cQ,\cQ_a\} &=& p_a~,\\
&\vdots&
\eea
The presence of the nilpotent scalar supercharge $\cQ$ is most important: it is the algebra of this charge that we can hope to translate to the lattice. The second piece of the algebra expresses the fact that the momentum is the $\cQ$-variation of something, which makes plausible the statement that the energy-momentum tensor and hence the entire action can be written in $\cQ$-exact form\footnote{Actually in the case of the four-dimensional $\cN=4$ there is an additional $\cQ$-closed term needed.}. Notice that an action written in such a $\cQ$-exact form is trivially invariant under the scalar supersymmetry, provided the latter remains nilpotent under discretization.

The rewriting of the supercharges in terms of twisted variables can be repeated for the fermions of the theory and yields a set of antisymmetric tensors $(\eta, \psi_a, \chi_{ab}, \ldots)$, which for the case of $Q=2^d$ matches the number of components of a real \KD field. This repackaging of the fermions of the theory into a \KD field is at the heart of how the discrete theory avoids fermion doubling as was shown by Becher, Joos and Rabin in the early days of lattice gauge theory \cite{Rabin:1981qj, Becher:1982ud}. 

It is important to recognize that the transformation to twisted variables corresponds to a simple change of variables in flat space -- one more suitable to discretization. A true topological field theory only results when the scalar charge is treated as a true BRST charge and attention is restricted to states annihilated by this charge. In the language of the supersymmetric parent theory such a restriction corresponds to a projection to the vacua of the theory. It is {\it not} employed in the lattice constructions we discuss in this paper.

\subsection{A warm up example: Twisted $\cN=2$ SYM in two dimensions}

We look at the twisted $\cN=2$ SYM in two dimensions as a warm up example. This theory satisfies our requirements for supersymmetric latticization: its R-symmetry possesses an $SO(2)$ subgroup corresponding to rotations of the two degenerate Majorana fermions into each other. Explicitly the theory can be written in twisted form as
\beq
S = \frac{1}{g^2} \cQ \int d^2x \Tr \left(\chi_{ab} \cF_{ab} + \eta [ \cDb_a,\cD_a ] - \frac{1}{2} \eta d \right)~.
\label{2daction}
\eeq
The degrees of freedom are just the twisted fermions $(\eta, \psi_a, \chi_{ab})$ previously described and a complex gauge field $\cA_a$. The latter is built from the usual gauge field and the two scalars present in the untwisted theory $\cA_a = A_a + i B_a$ with corresponding complexified field strength $\cF_{ab}$.

The complex covariant derivatives appearing in these expressions are defined by
\bea
\cD_a &=& \partial_a + \cA_a = \partial_a + A_a + iB_a~, \nn \\
\cDb_a &=& \partial_a + \cAb_a = \partial_a + A_a- i B_a~.
\eea
All fields take values in the adjoint representation of $U(N)$\footnote{The generators are taken to be {\it anti-hermitian} matrices satisfying $\Tr (T^aT^b)=-\delta^{ab}$.}. It should be noted that despite the appearance of a complexified connection and field strength, the theory possesses only the usual $U(N)$ gauge-invariance corresponding to the real part of the gauge field.

Notice that the original scalar fields transform as vectors under the original R-symmetry and hence become vectors under the twisted rotation group while the gauge fields are singlets under the R-symmetry and so remain vectors under twisted rotations. This structure makes possible the appearance of a complex gauge field in the twisted theory. 

The nilpotent transformations associated with $\cQ$ are given explicitly by
\bea
\cQ\; \cA_a &=& \psi_a \nn \\
\cQ\; \psi_a &=& 0 \nn \\
\cQ\; \cAb_a &=& 0 \nn \\
\cQ\; \chi_{ab} &=& -\cFb_{ab} \nn \\
\cQ\; \eta &=& d \nn \\
\cQ\; d&=&0
\eea

Performing the $\cQ$-variation and integrating out the auxiliary field $d$ yields
\beq
S = \frac{1}{g^2} \int d^2x \Tr \left(-\cFb_{ab}\cF_{ab} + \frac{1}{2}[ \cDb_a, \cD_a]^2 - \chi_{ab}\cD_{\left[a\right.}\psi_{\left.b\right]} - \eta \cDb_a\psi_a\right)~.
\label{twist_action}
\eeq

To untwist the theory and verify that indeed in flat space it just corresponds to the usual theory one can do a further integration by parts to produce 
\beq
S = \frac{1}{g^2} \int d^2x \Tr \left(-F^2_{ab} + 2B_a D_b D_b B_a - [B_a,B_b]^2 + L_F \right)~,
\eeq
where $F_{ab}$ is the usual Yang--Mills term. It is now clear that the imaginary parts of the gauge fields $B_a$ can now be given an interpretation as the scalar fields of the usual formulation. Similarly one can build spinors out of the twisted fermions and write the action in the manifestly Dirac form
\beq
L_F = \left(\begin{array}{cc}\chi_{12}&\frac{\eta}{2}\end{array}\right) \left(\begin{array}{cc}-D_2-iB_2&D_1+iB_1 \\
D_1-iB_1&D_2-iB_2\end{array}\right)
\left(\begin{array}{c}\psi_1 \\ 
\psi_2
\end{array}\right)~.
\eeq

\subsection{Discretization of the twisted $\cN=2$, $d=2$ theory}

\begin{figure}
\begin{center}
\includegraphics[width=0.5\textwidth]{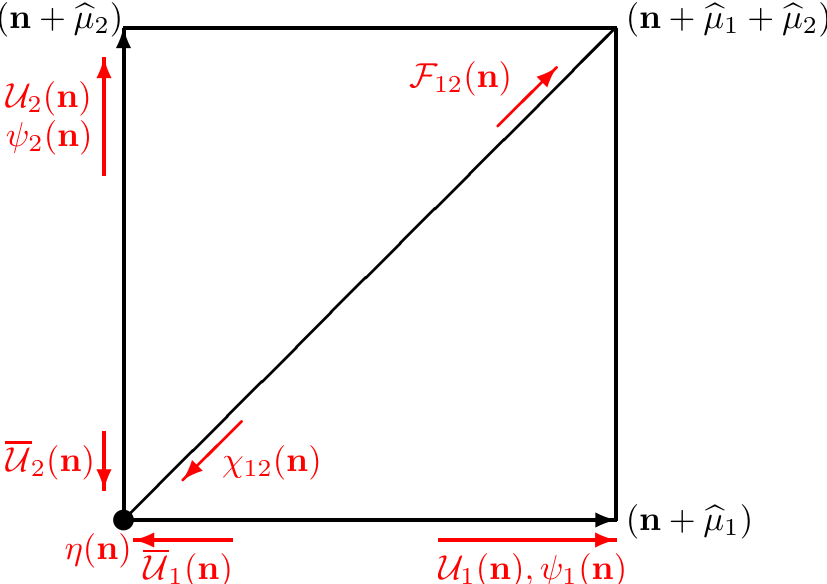}
\label{fig:2dlatt}
\end{center}
\caption{The unit cell of the two-dimensional $\cN=2$ lattice SYM with the orientation assignments for twisted fields.}
\end{figure}

The twisted theory described in the previous section may be discretized using the techniques developed in \cite{Catterall:2007kn, Damgaard:2007be, Damgaard:2008pa}. (Complex) gauge fields are represented as complexified Wilson gauge links $\cU_a(\vn)=e^{\cA_a(\vn)}$ living on links $(\vn, \vn+\hatbmu_a)$ of a lattice, which for the moment we can think of as hypercubic. These transform in the usual way under $U(N)$ lattice gauge transformations
\beq
\cU_a(\vn)\to G(\vn)\cU_a(\vn)G^\dagger(\vn + \hatbmu_a)~.
\eeq
Supersymmetric invariance then implies that $\psi_a(\vn)$ live on the same links as $\cU_a(\vn)$ and transform identically. The scalar fermion $\eta(\vn)$ is clearly most naturally associated with a site and transforms accordingly
\beq 
\eta(\vn)\to G(\vn)\eta(\vn)G^\dagger(\vn)~.
\eeq
The field $\chi_{ab}(\vn)$ is slightly more difficult. Naturally as a 2-form it should be associated with a plaquette. In practice we introduce diagonal links running through the center of the plaquette and choose $\chi_{ab}(\vn)$ to lie {\it with opposite orientation} along those diagonal links. This choice of orientation will be necessary to ensure gauge-invariance. Fig. 1 shows the unit cell of the resultant lattice theory.

To complete the discretization we need to describe how continuum derivatives are to be replaced by difference operators. A natural technology for accomplishing this in the case of adjoint fields was developed many years ago and yields expressions for the derivative operator applied to arbitrary lattice p-forms \cite{Aratyn:1984bd}. In the case discussed here we need just two derivatives given by the expressions
\bea
\cD^{(+)}_a f_b(\vn) &=& \cU_a(\vn)f_b(\vn+\hatbmu_a)-f_b(\vn)\cU_a(\vn+\hatbmu_b)~, \\
\cDb^{(-)}_a f_a(\vn) &=& f_a(\vn)\cUb_a(\vn)-\cUb_a(\vn-\hatbmu_a)f_a(\vn - \hatbmu_a)~.
\eea
The lattice field strength is then given by the gauged forward difference $\cF_{ab}(\vn) = \cD^{(+)}_a \cU_b(\vn)$ and is automatically antisymmetric in its indices. Furthermore, it transforms like a lattice 2-form and yields a gauge-invariant loop on the lattice when contracted with $\chi_{ab}(\vn)$. Similarly the covariant backward difference appearing in $\cDb^{(-)}_a \cU_a(\vn)$ transforms as a 0-form or site field and hence can be contracted with the site field $\eta(\vn)$ to yield a gauge-invariant expression.

This use of forward and backward difference operators guarantees that the solutions of the theory map one-to-one with the solutions of the continuum theory and hence fermion doubling problems are evaded \cite{Rabin:1981qj}. Indeed, by introducing a lattice with half the lattice spacing one can map this \KD fermion action into the action for staggered fermions \cite{Banks:1982iq}. Notice that, unlike the case of QCD, there is no rooting problem in this supersymmetric construction since the additional fermion degeneracy is already required by the continuum theory. The number of fermions of the twisted theory exactly fills out the components of a \KD field and corresponds to the taste degeneracy of (reduced) staggered fermions.

\subsection{Twisted $\cN=4$ SYM in three dimensions}

The twist of $\cN=4$ SYM in three dimensions\footnote{This twist of $\cN=4$, $d=3$ SYM is known as the Blau-Thompson twist \cite{Blau:1996bx}.} can be most succinctly written in the form where 
\bea
S &=&\frac{1}{g^2}Q \int d^3x \left(\chi_{ab}\cF_{ab} + \eta \left[\cDb_a,\cD_a\right] + \frac{1}{2}\eta d + B_{abc}\cDb_c \chi_{ab}\right)~.
\label{action1}
\eea
The fermions comprise a multiplet of p-form fields $(\eta, \psi_a, \chi_{ab}, \theta_{abc})$\footnote{It is common in the continuum literature to replace the 2- and 3-form fields in these expressions by their Hodge duals; a second vector $\hat{\psi}_a$ and scalar $\hat{\eta}$ see, for example \cite{Blau:1996bx}.} where in three dimensions $p=0, \cdots, 3$. This multiplet of twisted fermions corresponds to a single \KD field and here possesses eight single component fields as expected for a theory with $\cN=4$ supersymmetry in three dimensions.

The imaginary parts of the complex gauge field $\cA_a$ with $a=1, 2, 3$ appearing in this construction yield the three scalar fields of the conventional (untwisted) theory. Fields $d$ and $B_{abc}$ are auxiliaries introduced to render the scalar nilpotent supersymmetry $Q$ nilpotent off-shell. The latter acts on the twisted fields as follows
\bea
Q \cA_a &=& \psi_a \nn \\
Q \cAb_a &=& 0 \nn \\
Q \psi_a &=& 0 \nn \\
Q \chi_{ab} &=& \cFb_{ab} \\
Q \eta &=& d \nn \\
Q d &=& 0 \nn \\
Q B_{abc} &=& \theta_{abc} \nn \\
Q \theta_{abc} &=& 0\nn
\label{susy}
\eea
Notice that this construction differs slightly from the one discussed in \cite{Catterall:2010ng}. The fermion term involving a 3-form is here trivially rewritten as a $Q$-exact rather than $Q$-closed form. 

\begin{figure}
\begin{center}
\includegraphics[width=0.5\textwidth]{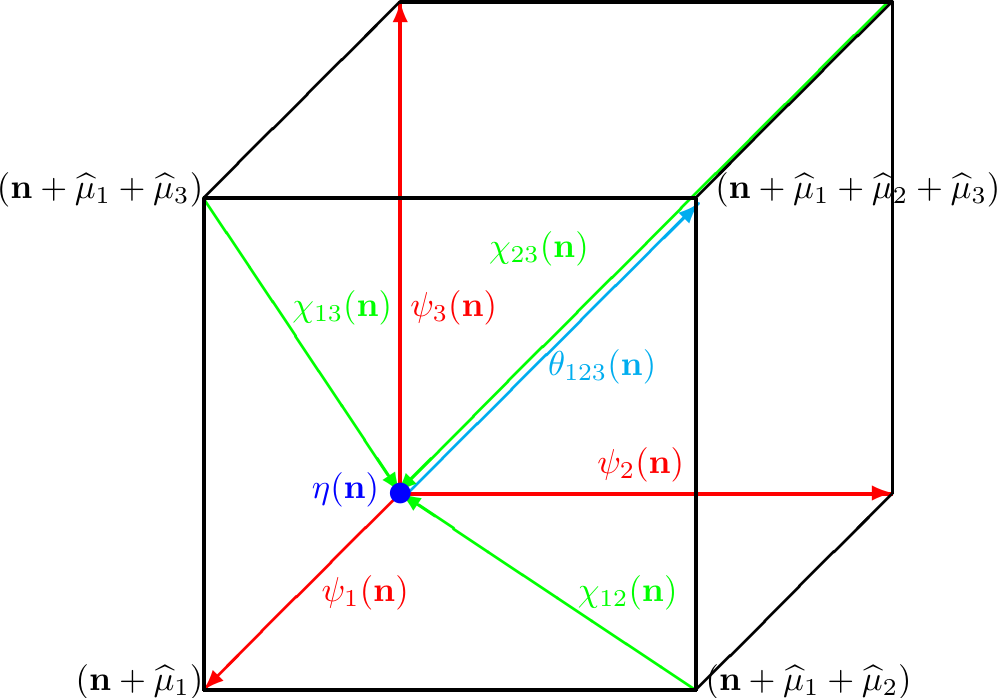}
\end{center}
\label{fig1}
\caption{The unit cell of the three-dimensional $\cN=4$ lattice SYM with the orientation assignments for twisted fermionic fields.}
\end{figure}

Doing the $Q$-variation, integrating out the field $d$ and using the Bianchi identity 
\beq
\epsilon_{abc}\cDb_c \cFb_{ab} = 0~,
\eeq
yields
\bea
\label{action}
S &=& \frac{1}{g^2}\int d^3 x~\Tr \Big(-\cFb_{ab}\cF_{ab} + \frac{1}{2}[ \cDb_a, \cD_a]^2 -\chi_{ab}\cD_{\left[a\right.}\psi_{\left.b\right]} \nn \\
&&- \psi_a\cDb_a\eta - \theta_{abc}\cDb_{\left[c \right.} \chi_{\left.ab\right]}\Big)~.
\eea
The terms appearing in the bosonic part of the action can then be written in the following form exposing the $B_a$ dependence explicitly
\bea
\cFb_{ab}\cF_{ab} &=& (F_{ab} - [B_a, B_b])(F_{ab} - [B_a, B_b]) + (D_{\left[a\right.}B_{\left.b\right]})
(D_{\left[a\right.}B_{\left.b\right]})~,\nn \\
\frac{1}{2}\left[\cDb_a, \cD_a\right]^2 &=& -2\left(D_a B_a\right)^2~,
\eea
where $F_{ab}$ and $D_a$ denote the usual field strength and covariant derivative depending on the real part of the connection $\cA_a$. 

\subsection{Discretization of the three-dimensional $\cN=4$ SYM theory}

The transition to the lattice from the continuum theory is similar to the case of the two-dimensional $\cN=2$ SYM theory. We replace the continuum complex gauge field $\cA_a(x)$ at every point by an appropriate complexified Wilson link $\cU_a(\vn)=e^{\cA_a(\vn)},~a = 1, 2, 3$. These lattice fields are taken to be associated with unit length vectors in the coordinate directions $\ba$ in an three-dimensional hypercubic lattice. By supersymmetry the fermion fields $\psi_a(\vn),~a = 1, 2, 3$ lie on the same oriented link as their bosonic superpartners running from $\vn \to \vn + \hatbmu_a$. In contrast the scalar fermion $\eta(\vn)$ is associated with the site $\vn$ of the lattice and the tensor fermions $\chi_{ab}(\vn),~a < b = 1, 2, 3$ with a set of diagonal face links running from $\vn + \hatbmu_a + \hatbmu_b \to \vn$. The final 3-form field $\theta_{abc}(\vn)$ is then naturally placed on the body diagonal running from $\vn \to \vn + \hatbmu_a + \hatbmu_b + \hatbmu_c$. The unit cell and fermionic field orientations of the three-dimensional theory is given in Fig. 2. The construction then posits that all link fields transform as bi-fundamental fields under gauge transformations
\bea
\eta(\vn) &\rightarrow& G(\vn) \eta(\vn) G^{\dagger}(\vn) \nn \\
\psi_m(\vn) &\rightarrow& G(\vn) \psi_m(\vn) G^{\dagger}(\vn + \hatbmu_m) \nn \\
\chi_{mn}(\vn) &\rightarrow& G(\vn + \hatbmu_m + \hatbmu_n) \chi_{mn}(\vn) G^{\dagger}(\vn)\\
\theta_{mnq}(\vn) &\rightarrow& G(\vn) \theta_{mnq}(\vn) G^{\dagger}(\vn + \hatbmu_m + \hatbmu_n + \hatbmu_q) \nn \\
\cU_m(\vn) &\rightarrow& G(\vn) \cU_m(\vn) G^{\dagger}(\vn + \hatbmu_m) \nn \\
\cUb_m(\vn) &\rightarrow& G(\vn + \hatbmu_m) \cUb_m(\vn) G^{\dagger}(\vn) \nn
\label{eq:gaugetrans-3d}
\eea  

The action of the lattice theory resembles to its continuum cousin with the one modification that the continuum field $\cA_a(x)$ is replaced with the Wilson link $\cU_m(\vn)$ and the lattice field strength being defined as $\cF_{mn}(\vn) = \cD_m^{(+)}\cU_n(\vn)$. Thus the supersymmetric and gauge-invariant lattice action is 
\bea
S &=& \cQ \sum_{\vn,m,n,q} \Tr \Big(\chi_{mn}(\vn)\cF_{mn}(\vn) + \eta(\vn) \cDb_m^{(-)}\cU_m(\vn) \nn \\
&&+ \frac{1}{2}\eta(\vn) d(\vn) + B_{mnq}(\vn)\cDb_q^{(+)}\chi_{mn}(\vn)\Big)~.
\eea

The covariant difference operators appearing in these expressions are defined by \cite{Damgaard:2008pa}
\bea
\cD_m^{(-)}f_m(\vn) &=& \cU_m(\vn)f_m(\vn) - f_m(\vn - \hatbmu_m)\cU_m(\vn - \hatbmu_m)\\
\cD_m^{(+)}f_n(\vn) &=& \cU_m(\vn)f_n(\vn + \hatbmu_m) - f_n(\vn)\cU_m(\vn + \hatbmu_n)\\
\cDb_m^{(-)}f_m(\vn) &=& f_m(\vn)\cUb_m(\vn) - \cUb_m(\vn-\hatbmu_m)f_m(\vn-\hatbmu_m)\\
\cDb_m^{(+)}f_{nq}(\vn) &=& f_{nq}(\vn+\hatbmu_m)\cUb_m(\vn) - \cUb_m(\vn+\hatbmu_n+\hatbmu_q)f_{nq}(\vn)
\eea
These expressions are determined by the twin requirements that they reduce to the corresponding continuum results for the adjoint covariant derivative in the naive continuum limit $\cU_m \rightarrow {\mathbb I}_N + \cA_m$ and that they transform under gauge transformations like the corresponding lattice link field carrying the same indices. This allows the terms in the action to correspond to gauge-invariant closed loops on the lattice. 

Upon following the prescription \cite{Damgaard:2008pa} for lattice covariant derivatives, we write down the lattice action in terms of the link fields $\cU_m(\vn)$ and $\cUb_m(\vn)$
\bea
\label{eq:4action}
S &=& \frac{1}{g^2}\sum_{\vn,m,n,q} \Tr \Big(-\cFb_{mn}(\vn)\cF_{mn}(\vn) + \frac{1}{2}\Big(\cDb_m^{(-)}\cU_m(\vn)\Big)^2 \nn \\
&&-\chi_{mn}(\vn)\cD_{[m}^{(+)}\psi_{n]}(\vn) - \eta(\vn) \cDb_m^{(-)}\psi_m(\vn) - \theta_{mnq}(\vn)\cDb_{[q}^{(+)}\chi_{mn]}(\vn)\Big)~.~~~~~~
\eea

The bosonic part of the action is
\bea
S_B &=& \frac{1}{g^2}\sum_{\vn,m,n} \Tr \Big[-\Big(\overline{\cD_m^{(+)}\cU_n(\vn)}\Big)\Big(\cD^{(+)}_m \cU_n(\vn)\Big) + \frac{1}{2} \Big(\cDb^{(-)}_{m} \cU_{m}(\vn)\Big)^2\Big]\nn \\
&=& \frac{1}{g^2}\sum_{\vn,m,n} \Tr \Big[\Big(\cUb_n(\vn + \hatbmu_m)\cUb_m(\vn) - \cUb_m(\vn + \hatbmu_n)\cUb_n(\vn)\Big)\nn \\
&& \times \Big(\cU_{m}(\vn)\cU_{n}(\vn + \hatbmu_m) - \cU_n(\vn)\cU_m(\vn + \hatbmu_n)\Big)\nn \\
&&+ \frac{1}{2} \Big(\cU_m(\vn)\cUb_m(\vn) - \cUb_m(\vn - \hatbmu_m)\cU_m(\vn - \hatbmu_m)\Big)^2\Big]~,
\eea
and the fermionic part
\bea
S_F &=& -\frac{1}{g^2}\sum_{\vn,m,n,q,r,e,f} \Tr \Big\{\frac{1}{2}(\delta_{mq}\delta_{nr} - \delta_{mr}\delta_{nq}) \nn \\
&& \times \Big[\chi_{mn}(\vn)\Big(\cU_q(\vn)\psi_r(\vn + \hatbmu_q) - \psi_r(\vn)\cU_q(\vn + \hatbmu_r)\Big)\Big] \nn \\
&&+ \eta(\vn)\Big(\psi_m(\vn)\cUb_m(\vn) - \cUb_m(\vn - \hatbmu_m)\psi_m(\vn - \hatbmu_m)\Big)\nn \\
&&+\frac{1}{3}(\delta_{mr}\delta_{ne}\delta_{qf} + \delta_{qr}\delta_{me}\delta_{nf} + \delta_{nr}\delta_{qe}\delta_{mf})\nn \\
&&\times \theta_{ref}(\vn)\Big(\chi_{re}(\vn+\hatbmu_f)\cUb_f(\vn) - \cUb_f(\vn+\hatbmu_r+\hatbmu_e)\chi_{re}(\vn)\Big)\Big\}~.
\eea

It is easy to see that each term in the lattice action forms a gauge-invariant loop on the lattice.

\subsection{Twisted $\cN=4$ SYM in four dimensions}

In four dimensions the constraint that the target theory possess 16 supercharges singles out a single theory for which this construction can be undertaken -- the $\cN=4$ SYM. 

The continuum twist of $\cN=4$ that is the starting point of the twisted lattice construction was first written down by Marcus in 1995 \cite{Marcus:1995mq} although it now plays an important role in the Geometric-Langlands program and is, hence, sometimes called the GL-twist \cite{Kapustin:2006pk}. This four-dimensional twisted theory is most compactly expressed as the dimensional reduction of a five-dimensional theory in which the ten (one gauge field and six scalars) bosonic fields are realized as the components of a complexified five-dimensional gauge field while the 16 twisted fermions naturally span one of the two \KD fields needed in five dimensions. Remarkably, the action of this theory contains a $\cQ$-exact piece of precisely the same form as the two dimensional theory given in (\ref{2daction}) provided one extends the field labels to run now from one to five. In addition, the Marcus twist requires a new $\cQ$-closed term, which was not possible in the two-dimensional theory. 
\beq
S_{\rm closed} = -\frac{1}{8}\int \Tr \epsilon_{mnpqr}\chi_{qr}\cDb_p\chi_{mn}\label{closed}~.
\eeq
The supersymmetric invariance of this term then relies on the Bianchi identity 
\beq
\epsilon_{mnpqr}\cDb_p\cFb_{qr}=0~. 
\eeq

\subsection{Discretization of the four-dimensional $\cN=4$ SYM theory}

In two and three dimensions we were able to accommodate the bosonic fields of the theory in a natural way by assigning them to the links of a hypercubic lattice. For the $\cQ=16$ theory this is not possible; the theory can be parametrized in terms of five complex gauge fields in the continuum. We are thus motivated to search for a four-dimensional lattice with five basis vectors $\hatbmu_a$, $a=1,\cdots, 5$. One simple solution is to use a hypercubic lattice with an additional body diagonal
\bea
\label{eq:mu-vectores}
\hatbmu_1 &=& (1, 0, 0, 0)\nn \\
\hatbmu_2 &=& (0, 1, 0, 0)\nn \\
\hatbmu_3 &=& (0, 0, 1, 0) \\
\hatbmu_4 &=& (0, 0, 0, 1)\nn \\
\hatbmu_5 &=& (-1, -1, -1, -1)\nn
\eea
The field $\cU_5$ is then placed on the body diagonal link. Actually, we will indeed utilize such a hypercubic lattice when building the C++ data structure needed to code the resulting theory. Notice that the basis vectors sum to zero, consistent with the use of such a linearly dependent basis.
 
However, it should also be clear that a more symmetrical choice is possible in which the five basis vectors are entirely equivalent and the lattice theory possesses a large point group symmetry $S^5$ corresponding to permutations of the set of basis vectors. Such a discrete structure exists in four dimensions: it is called the $A_4^*$ lattice. It is constructed from the set of five basis vectors $\hatbe_a$ pointing from the center of a four-dimensional equilateral simplex out to its vertices together with their inverses $-\hatbe_a$. It is the four-dimensional analog of the two-dimensional triangular lattice. 
A specific basis for the $A_4^*$ lattice is given in the form of five lattice vectors
\bea
\hatbe_1 &=&  \Big(\frac{1}{\sqrt{2}}, \frac{1}{\sqrt{6}}, \frac{1}{\sqrt{12}}, \frac{1}{\sqrt{20}}\Big)\\
\hatbe_2 &=& \Big(-\frac{1}{\sqrt{2}}, \frac{1}{\sqrt{6}}, \frac{1}{\sqrt{12}}, \frac{1}{\sqrt{20}}\Big)\\
\hatbe_3 &=& \Big(0, -\frac{2}{\sqrt{6}}, \frac{1}{\sqrt{12}}, \frac{1}{\sqrt{20}}\Big)\\
\hatbe_4 &=& \Big(0, 0, -\frac{3}{\sqrt{12}}, \frac{1}{\sqrt{20}}\Big)\\
\hatbe_5 &=& \Big(0, 0, 0, -\frac{4}{\sqrt{20}}\Big)
\eea
The basis vectors satisfy the relations 
\beq
\sum_{m=1}^{5} \hatbe_m = 0;~\hatbe_m \cdot \hatbe_n = \Big(\delta_{mn} - \frac{1}{5}\Big);~\sum_{m=1}^{5}(\hatbe_m)_{\mu}(\hatbe_m)_{\nu} = \delta_{\mu \nu};~\mu, \nu = 1,\cdots,4.
\eeq
Notice that $S^5$ is a subgroup of the twisted rotation symmetry group $SO(4)^\prime$. Furthermore, the lattice fields transform in reducible representations of this discrete group - for example, the vector $\cU_a$ decomposes into a four component vector $\cU_\mu$ and a scalar field $\phi=\sum_a \cU_a$ under $S^5$ and hence also under $SO(4)^\prime$ in the continuum limit. Invariance of the lattice theory with respect to $S^5$ then guarantees that the lattice theory will inherit full invariance under twisted rotations as the lattice spacing is sent to zero.

Complexified Wilson gauge link variables $\cU_a$ are then placed on these links together with their $\cQ$-superpartners $\psi_a$. The ten twisted fermions $\chi_{ab}$ are associated with additional diagonal links $\hatbe_a+\hatbe_b$ with $a>b$ while a single fermion $\eta$ is placed at each lattice site. 

We can connect the basis vectors of the hypercubic lattice and the $A_4^*$ lattice through a set of linear transformations - see \cite{Unsal:2006qp, Catterall:2011pd}. The integer-valued hypercubic lattice site vector $\vn$ can be related to the physical location in space-time using the $A_4^*$ basis vectors $\hatbe_a$
\beq
\bR = a \sum_{\nu =1}^4 (\mu_{\nu} \cdot \vn)\hatbe_{\nu} = a \sum_{\nu =1}^{4}n_{\nu}\hatbe_{\nu}~,
\eeq
where $a$ is the lattice spacing. On using the fact that $\sum_{m}\hatbe_m = 0$, we can show that a small lattice displacement of the form $d\vn = \hatbmu_m$ corresponds to a space-time translation by $(a\hatbe_m)$:
\beq
d\bR = a \sum_{\nu =1}^{4}(\mu_{\nu} \cdot d \vn)\hatbe_{\nu} = a \sum_{\nu =1}^{4} (\hatbmu_{\nu} \cdot \hatbmu_m)\hatbe_{\nu} = a \hatbe_m~.
\eeq
The lattice action corresponds to a discretization of the Marcus twist on this $A_4^*$ lattice and can be represented as a set of traced closed bosonic and fermionic loops. It is invariant under the exact $\cQ$ scalar supersymmetry, lattice gauge transformations and a global permutation (point group) symmetry $S^5$, and can be proven free of fermion doubling problems as discussed before. The $\cQ$-exact part of the lattice action is again given by (\ref{twist_action}) with the indices $\mu,\nu$ now labeling the five basis vectors of $A_4^*$ or equivalently its hypercubic cousin.

Finally, it is important to note that while the true lattice in space-time is this rather complicated looking $A_4^*$ structure, we can represent all of the lattice fields in our theory by giving only their coordinates on the abstract hypercubic lattice. Indeed, since the lattice action only depends on the structure of the hypercubic lattice we will not need the explicit coordinates of the $A_4^*$ lattice to generate Monte Carlo configurations during the simulation. The explicit mapping of hypercubic coordinates to space-time coordinates in the $A_4^*$ lattice is only needed when, for example, we want to compute spatially dependent objects such as correlation functions of fields. In this case we should compute distances relative to the underlying $A_4^*$ lattice {\it not} its hypercubic partner.
 
While the supersymmetric invariance of the $\cQ$-exact term is manifest in the lattice theory it is not immediately clear how to discretize the continuum $\cQ$-closed term. Remarkably, it is possible to discretize (\ref{closed}) in such a way that it is indeed exactly invariant under the twisted supersymmetry: 
\beq
S_{\rm closed} = -\frac{1}{8} \sum_{\vn,m,n,p,q,r} \Tr \epsilon_{mnpqr}\chi_{qr}(\vn+\hatbmu_m+\hatbmu_n+\hatbmu_p) \cDb^{(-)}_p \chi_{mn}(\vn+\hatbmu_p)~,
\eeq
which can be seen to be supersymmetric since the lattice field strength satisfies an exact Bianchi identity \cite{Aratyn:1984bd}
\beq
\epsilon_{mnpqr}\cDb^{(+)}_p\cFb_{qr}=0~.
\eeq

\section{Simulating the SYM theories: Algorithms}
\label{sec:simulating-sym-alg}

Although the fields entering into these twisted descriptions appear somewhat different to the usual fields used in QCD the basic algorithms we use to simulate them are borrowed directly from lattice QCD; namely we integrate out the fermions to produce a Pfaffian which is in turn represented by the square root of a determinant\footnote{Of course this ignores a possible sign ambiguity. We return to this issue later when we discuss whether the phase quenched simulations we use suffer from a sign problem.} and can be simulated using the usual RHMC algorithm \cite{Clark:2006wq}.

If we denote the set of twisted fermions by the field $\Psi=(\eta, \psi_\mu, \chi_{\mu\nu})$ we first introduce a corresponding pseudo-fermion field $\Phi$ with action
\beq
S_{\rm PF}=\Phi^\dagger (M^\dagger M)^{-\frac{1}{4}} \Phi~,
\label{pseudo}
\eeq
where $M=M(\cU,\cU^\dagger)$ is the antisymmetric twisted lattice fermion operator given, for example, in (\ref{eq:4action})\footnote{The antisymmetry
is guaranteed if the fermion action is rewritten as the sum of the original terms plus their lattice transposes.}. 

Integrating over the fields $\Phi$ will then yield (up to a possible phase) the Pfaffian of the operator $M(\cU, \cU^\dagger)$ as required. The fractional power is approximated by the partial fraction expansion
\beq
\frac{1}{(M^\dagger M)^{\frac{1}{4}}} = \alpha_0 + \sum_{i=1}^P\frac{\alpha_i}{M^\dagger M + \beta_i}~,
\label{partial}
\eeq
where the coefficients $\{\alpha_i, \beta_i\}$ are evaluated offline using the Remez algorithm to minimize the error in some interval $(\epsilon, A)$. Typically we have used $P=15$ which yields a fractional error of $0.00001$ for the interval $0.0000001 \to 1000.0$, which conservatively covers the range we are interested in.

Following the standard procedure, we introduce momenta $(p_\cU, p_F)$ conjugate to the coordinates $(\cU, \Phi)$ and evolve the coupled system using a discrete time leapfrog algorithm according to the classical Hamiltonian 
\beq
\label{eq:class-Hamilt}
H = S_B + S_{\rm PF} + p_\cU\bar{p}_\cU + p_\Phi\bar{p}_\Phi~.
\eeq
Notice that the bosonic action\footnote{From now on we interchangeably use $\bx$ and $\vn$ to denote the lattice site.}  
\beq
S_B=\sum_{\vx,m,n} \Tr \Big[-\Big(\overline{\cD_m^{(+)}\cU_n(\vx)}\Big)\Big(\cD^{(+)}_m \cU_n(\vx)\Big) + \frac{1}{2} \Big(\cDb^{(-)}_{m} \cU_{m}(\vx)\Big)^2\Big]~,
\eeq 
is real, positive semi-definite in all these theories.

One step of the discrete time update is given by
\bea
\delta p_\cU&=&\frac{\delta t}{2}\bar{f}_\cU\\
\delta p_\Phi&=&\frac{\delta t}{2}\bar{f}_\Phi\\
\delta\cU&=&\left(e^{\delta t p_\cU}-I\right)\cU\\
\delta\Phi&=&\delta t p_{\Phi}\\
\delta p_\cU&=&\frac{\delta t}{2}\bar{f}_\cU\\
\delta p_\Phi&=&\frac{\delta t}{2}\bar{f}_\Phi
\eea
where the forces $f_\cU$ and $f_\Phi$ are given by
\bea
f_\cU&=&-\frac{\delta S}{\delta \cU}\\
f_\Phi&=&-\frac{\delta S}{\delta \Phi}
\eea
and the bar denotes complex conjugation. Using the partial fraction expansion given in (\ref{partial}) the fermionic contributions to these forces take the form
\bea
f^{fermionic}_\cU &=& \sum_{i=1}^P\alpha_i\left[\bar{t}_i\frac{\delta M}{\delta\cU}s_i + \overline{\left(\bar{t}_i\frac{\delta M}{\delta\cUb}s_i\right)}
\right]\\
f^{fermionic}_\Phi &=& -\alpha_0\bar{\Phi}-\sum_{i=1}^P\alpha_i\bar{s}_i\\
\eea
where 
\bea
(M^\dagger M+\beta_i)s_i &=& \Phi\\
t_i &=& Ms_i
\eea
The latter set of sparse linear equations is solved using a multi-mass conjugate gradient (MCG) solver \cite{Jegerlehner:1996pm}, which allows for the simultaneous solution of all $P$ systems in a single CG solve.
 
At the end of one such classical trajectory the final configuration is subjected to a standard Metropolis test based on the Hamiltonian $H$. The symplectic and reversible nature of the discrete time update is then sufficient to allow for detailed balance to be satisfied and hence expectation values are independent of $\delta t$. After each such trajectory the momenta are refreshed from the appropriate Gaussian distribution as determined by $H$, which renders the simulation ergodic. 

The fermionic contribution to the forces are shown below
\bea
f^{fermionic}_{\cU_m} &=& \frac{\partial S_{pf}}{\partial \cU_m} = \sum_{i=1}^{P} \alpha_i F^{\dagger} \frac{-1}{(M^{\dagger}M + \beta_i)^2}\frac{\partial}{\partial \cU_m} (M^{\dagger}M) F \nn \\
&=& -\sum_{i=1}^{P} \alpha_i \Big(\frac{F}{(M^{\dagger}M + \beta_i)}\Big)^{\dagger} \frac{\partial}{\partial \cU_m} (M^{\dagger}M) \Big(\frac{F}{(M^{\dagger}M + \beta_i)}\Big) \nn \\
&=& -\sum_{i=1}^{P} \alpha_i \Big(\frac{F}{(M^{\dagger}M + \beta_i)}\Big)^{\dagger} \Big(M^{\dagger}\frac{\partial M}{\partial \cU_m} + \frac{\partial M^{\dagger}}{\partial \cU_m}M\Big) \Big(\frac{F}{(M^{\dagger}M + \beta_i)}\Big) \nn \\
&=& -\sum_{i=1}^{P} \alpha_i \Big[\Big(M \frac{F}{(M^{\dagger}M + \beta_i)}\Big)^{\dagger} \frac{\partial M}{\partial \cU_m}\Big(\frac{F}{(M^{\dagger}M + \beta_i)}\Big) \nn \\
&&+ \Big(\frac{F}{(M^{\dagger}M + \beta_i)}\Big)^{\dagger} \frac{\partial M^{\dagger}}{\partial \cU_m} \Big(M \frac{F}{(M^{\dagger}M + \beta_i)}\Big)\Big] \nn \\
&=& -\sum_{i=1}^{P} \alpha_i \Big[t_i^{\dagger} \frac{\partial M}{\partial \cU_m}s_i + s_i^{\dagger} \frac{\partial M^{\dagger}}{\partial \cU_m} t_i\Big]~.
\eea
\bea
f^{fermionic}_{F} &=& \frac{\partial S_{pf}}{\partial F}\nn \\
&=& \alpha_0\frac{\partial}{\partial F} (F^{\dagger}F) + \sum_{i=1}^{P} \alpha_i \frac{\partial}{\partial F} \Big(F^{\dagger}\Big[(M^{\dagger}M + \beta_i)^{-1}F\Big]\Big) \nn \\
&=&\alpha_0 F^{\dagger} + \sum_{i=1}^{P} \alpha_i s_i^{\dagger}~.
\eea

\section{Overall structure of the C++ code}
\label{sec:overall-structure}

Typically the bosons lie on the usual nearest neighbor links of a hyeprcubic lattice while the fermions occupy both these links and additional site, face and body diagonal links. In the case of $\cN=4$ in four dimensions we have to augment the set of boson links with one additional gauge field associated with the body diagonal link of the hypercube. We introduce the {\tt Lattice\_Vector} class to store the coordinates of the lattice sites and also the vector between sites. Such lattice vectors can be added or subtracted by overloading the `$+$' or `$-$' operators. These operations also respect the lattice boundary conditions. Associated with this class is a general function {\tt loop\_over\_lattice(x sites)} that implements a loop over all lattice sites indexed by their coordinate vector; thus a simple loop looks like
\bec 
{\tt while(loop\_over\_lattice(x,sites)){....}}
\eec 

The bosonic and pseudo fermionic fields are stored in various objects which are indexed via their lattice site vector and whose type corresponds directly to the tensor structure of the associated continuum field so that one finds C++ classes labeled {\tt Site\_Field}, {\tt Link\_Field}, {\tt Plaq\_Field}, {\tt Body\_Field} etc. in the header file {\tt utilities.h}. (We provide the list of C++ files that goes into the code in \ref{sec:file-list}.) The full \KD field is contained in the class {\tt Twist\_Fermion} while the {\tt Gauge\_Field} class contains the complexified Wilson gauge link. All these objects are in turn built from objects of type {\tt Umatrix} corresponding to complex {\tt NCOLOR x NCOLOR} matrices. Simple arithmetric operations which overload the usual arithmetic operations are defined for manipulating these objects.
\begin{figure}
\begin{center}
\includegraphics[width=0.8\textwidth]{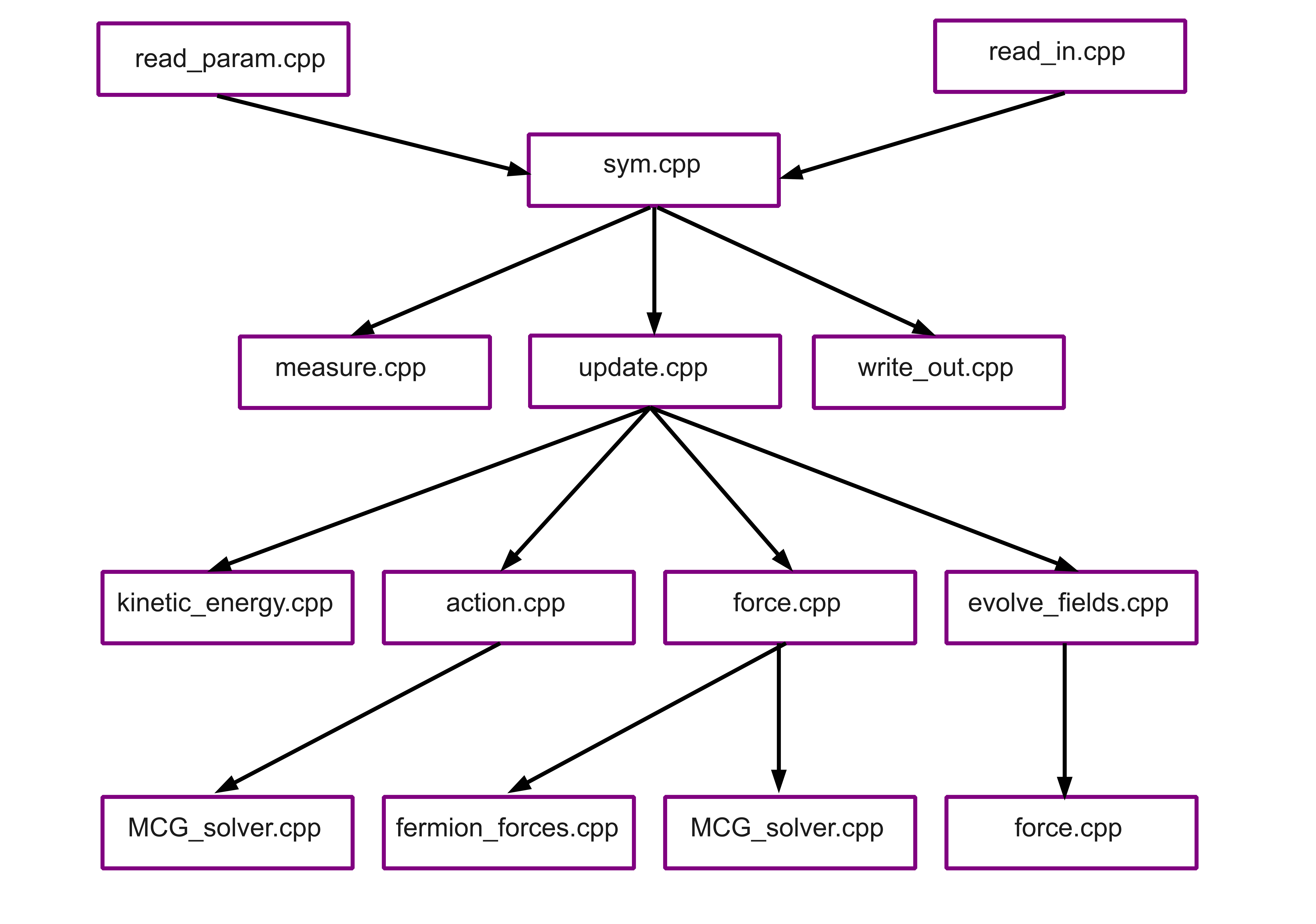}
\end{center}
\label{fig:structure}
\caption{The organizational structure of the C++ code that generates and measures field configurations.}
\end{figure}

Let us briefly describe how the code works. The general organizational structure of the code is given in Fig. 3. We begin with {\tt sym.cpp}. It reads the input parameters such as number of sweeps ({\tt SWEEPS}), number of thermalization steps ({\tt THERM}), gap in measurements ({\tt GAP}), the `t Hooft coupling ({\tt LAMBDA}), etc., using functions contained in the file {\tt read\_param.cpp}. It can also read in previously generated field configurations using {\tt read\_in.cpp}.
{\ }\\

The code {\tt sym.cpp} performs three major tasks: 
\begin{itemize}
\item[1.] Generates new configurations using a rational hybrid Monte Carlo (RHMC) algorithm. This is accomplished by calling the function {\tt update(U,F)} contained in {\tt update.cpp}.
\item[2.] Saves the current field configuration after some number of Monte Carlo sweeps (using the functions in {\tt write\_out.cpp}).
\item[3.] Measures the observables in the theory. This is done by function calls within {\tt measure.cpp}.
\end{itemize}  

Let us focus on the task of updating field configurations first. After reading the initial parameters and field configurations {\tt update()} is called. Here we refresh the momenta ${\tt p\_U}$ and ${\tt p\_F}$ (using a Gaussian distribution) and then go to {\tt kinetic\_energy.cpp} to compute the kinetic energy: 
\bec
{\tt Adj(p\_U)*p\_U + Cjg(p\_F)*p\_F}.
\eec
Compare this with the first two terms in the classical Hamiltonian (\ref{eq:class-Hamilt}):
\bec
$\overline{p}_{\cU}p_{\cU} +  \overline{p}_{\Phi}p_{\Phi}$~.
\eec
After computing kinetic energy the boson and pseudo-fermion actions (\ref{eq:class-Hamilt}) are computed with a call to the function {\tt action()}. 

The computation of the bosonic action $S_B$ is straightforward. In the code it is accomplished with the line
\bec
{\tt KAPPA*[0.5*Tr(DmuUmu*DmuUmu) + 2.0*Tr(Fmunu*Adj(Fmunu))] }~.
\eec 
Here {\tt KAPPA} is the dimensionless lattice coupling. It is defined in {\tt read\_param.cpp} and depends on the number of dimensions ({\tt D}), size of the lattice ({\tt LX}, {\tt LY}, {\tt LZ}, {\tt T}) and number of colors ({\tt NCOLOR}).
 
The code associated with spefcific terms in the bosonic action can easily be identified with its analytic expression. We have
\bec
{\tt DmuUmu(x)} $\rightarrow$ {\tt Umu(x)*Udagmu(x)-Udagmu(x-e\_mu)*Umu(x-e\_mu)}~, \\
{\tt Fmunu(x)} $\rightarrow$ {\tt Umu(x)*Unu(x+e\_mu)-Unu(x)*Umu(x+e\_nu)}~.
\eec

The code used to compute the fermionic part of the action is given by
\bec
{\tt S\_F =  ampdeg*(Cjg(F)*F) + $\sum_{\tt n=0}^{\tt DEGREE}$ amp[n]*(Cjg(F)*sol[n])}~,
\eec
where {\tt n} runs from ${\tt 0}$ to {\tt DEGREE} (which is equal to number of terms in the Remez approximation $P$), {\tt ampdeg} corresponds to $\alpha_0$, {\tt F} the twisted pseudo-fermion $F$, {\tt Cjg(F)} is $F^{\dagger}$, {\tt amp[n]} is $\alpha_i$  and {\tt sol[n]} corresponds to $s_i \equiv (M^{\dagger}M + \beta_i)^{-1}F$.

Again one should compare this code with the form of the pseudo-fermion action
\bec
$S_{pf} = \alpha_0 F^{\dagger}F + \sum_{i=1}^{P} \alpha_i F^{\dagger}\Big[(M^{\dagger}M + \beta_i)^{-1}F\Big]$~.
\eec

We invoke a multi-mass conjugate gradient solver {\tt MCG\_solver()} given in {\tt MCG\_solver.cpp} to help compute the terms needed in the fermionic action. The MCG solver can return the solutions to $(M^{\dagger}M + \beta_i) s_i = F$ for all shifts $\beta_i$.

Once the Hamiltonian is computed we evolve the fields along a classical trajectory. This is handled by the function {\tt evolve\_fields}. The evolution of the fields and momenta is achieved through a leapfrog algorithm. In the first half step we have
\bec
\begin{tabular}{ r c l }
{\tt p\_Umu} & $\rightarrow$ & {\tt p\_Umu + 0.5*DT*f\_Umu} \\
{\tt p\_F} & $\rightarrow$ & {\tt p\_F + 0.5*DT*f\_F} \\
{\tt Umu} & $\rightarrow$ & {\tt Umu + exp(DT*p\_Umu)} \\
{\tt F} & $\rightarrow$ & {\tt F + DT*p\_F}
\end{tabular}
\eec

Immediately after computing the change in fields ({\tt Umu} and {\tt F}) and momenta ({\tt p\_Umu} and {\tt p\_F}), we update the forces by calling {\tt force()}. The bosonic force contribution to {\tt f\_Umu} is given by
\bec
\begin{tabular}{ r c l }
{\tt f\_Umu(x)} & $\rightarrow$ & {\tt f\_Umu(x)+Umu(x)*Udagmu(x)*DmuUmu(x)}\\
                &               & {\tt-Umu(x)*DmuUmu(x+e\_mu)*Udagmu(x)}\\ 
                &               & {\tt+2.0*Umu(x)*Unu(x+e\_mu)*Adj(Fmunu(x))}\\
                &               & {\tt-2.0*Umu(x)*Adj(Fmunu(x-e\_nu))*Unu(x-e\_nu)}
\end{tabular}
\eec

The computation of the fermionic force {\tt f\_F} requires first a call to the MCG solver ${\tt MCG\_solver()}$. We find 
\bec
{\tt f\_F = -ampdeg*Cjg(F) - $\sum_{\tt n=0}^{\tt DEGREE}$ amp[n]*Cjg(sol[n])}~.
\eec
Once we have this solution an additional contribution to the gauge force coming from the pseudo-fermions is gotten by a call to the function {\tt fermion\_forces()}. Each fermionic term in the action yields a contribution. We provide a part of this code in Fig. 4. 
\begin{figure}
\begin{center}
\lstset{basicstyle=\footnotesize}
\fbox{
\begin{lstlisting}[language=C++]
#include "fermion_forces.h"

void fermion_forces(const Gauge_Field &U, Gauge_Field &f_U, 
	const Twist_Fermion &s, const Twist_Fermion &p)
{
  Lattice_Vector x, e_mu;
  int sites, mu, a, b;
  Umatrix tmp;
  Gauge_Field Udag;

  Udag=Adj(U);
  f_U=Gauge_Field();
  //contribution to f_U from psi_muDb_mu(U)eta term
    sites=0;
    while(loop_over_lattice(x,sites))
    {
      for(mu=0;mu<NUMLINK;mu++)
      {e_mu=Lattice_Vector(mu);
      tmp=Umatrix();
        for(a=0;a<NUMGEN;a++)
        {
          for(b=0;b<NUMGEN;b++)
          {tmp=tmp+conjug(p.getS().get(x).get(a))*s.getL().get(x,mu).get(b)
          *Lambda[a]*Lambda[b]*Udag.get(x,mu)-conjug(p.getS().get(x+e_mu).get(a))
          *BC(x,e_mu)*s.getL().get(x,mu).get(b)*Lambda[b]*Lambda[a]*Udag.get(x,mu);}
        }
        f_U.set(x,mu,f_U.get(x,mu)-0.5*Adj(tmp));}
    }
    sites=0;
    while(loop_over_lattice(x,sites))
    {
      for(mu=0;mu<NUMLINK;mu++)
      {e_mu=Lattice_Vector(mu);
      tmp=Umatrix();
        for(a=0;a<NUMGEN;a++)
        {
          for(b=0;b<NUMGEN;b++)
          {tmp=tmp+conjug(p.getL().get(x,mu).get(a))*s.getS().get(x+e_mu).get(b)
          *BC(x,e_mu)*Lambda[a]*Lambda[b]*Udag.get(x,mu)-
          conjug(p.getL().get(x,mu).get(a))*s.getS().get(x).get(b)
          *Lambda[b]*Lambda[a]*Udag.get(x,mu);}
        }
        f_U.set(x,mu,f_U.get(x,mu)-0.5*Adj(tmp));}
    }
    sites=0;
    while(loop_over_lattice(x,sites))
    {for(mu=0;mu<NUMLINK;mu++){f_U.set(x,mu,-1.0*Adj(f_U.get(x,mu)));}}
return;
}
\end{lstlisting}}
\end{center}
\caption{A part of the C++ code to compute the fermion force contribution.}
\label{fig:code}
\end{figure}
In the second half step of the leapfrog algorithm the momenta {\tt p\_U} and {\tt p\_F} are again updated with the new forces. These final forces are then saved for the next iteration.

In practice, it is important to use a multi-time step integrator for this evolution \cite{Sexton:1992nu}. In this case while the fermions are evolved with a time step of {\tt DT}, the bosons are integrated with the time step {\tt DT/MSTEP}. Provided the boson force is substantially larger than the fermionic contribution this can result in fewer costly fermion inversions for a fixed acceptance rate. In practice the parameter {\tt MSTEPS} can be tuned to optimize the update - typically {\tt MSTEPS=10}.

Finally, control returns to {\tt update()} and the updated Hamiltonian {\tt H\_new} is computed. A simple Metropolis test is used to accept or reject the field configuration at the end of the trajectory.

\subsection{Site, Link and Plaquette type operators}

The bosonic and fermionic fields, and the covariant difference operators living on the hypercubic lattice are associated with various geometric structures such as sites, links and plaquettes. They are implemented in the code using various user defined C++ classes: {\tt Site\_Field}, {\tt Link\_Field}, {\tt Plaq\_Field}, {\tt Body\_Field}, etc. They are constructed such that they can take values in $U(N)$ or $SU(N)$. They make appearances in the code in many ways and we summarize their general structure in the table below:
\bec
\begin{tabular}{| c | c | c | c |}\hline
{\tt Site\_Field} & $S(\vx)$ & $\cDb^{(-)}_\mu L_\mu(\vx)$ & \\ \hline 
{\tt Link\_Field} & $L_\mu(\vx)$ & $\cDb^{(+)}_\mu S(\vx)$ & $\cD^{(-)}_\nu P_{\mu \nu}(\vx)$ \\ \hline
{\tt Plaquette\_Field} & $P_{\mu \nu}(\vx)$ & $\cD^{(+)}_\mu L_\nu(\vx)$ & $\cDb^{(-)}_\rho B_{\mu \nu \rho}(\vx)$ \\ \hline
{\tt Body\_Field} & $B_{\rho \mu \nu}(\vx)$ & $\cDb^{(+)}_\rho P_{\mu \nu}(\vx)$ & \\
\hline
\end{tabular}
\eec

As an instructive example let us look at the coding details of the {\tt Link\_Field} class. In Fig. 5 we show how the {\tt Link\_Field} class is defined along with overloading of basic operators such as `+' and `-'.
\begin{figure}
\begin{center}
\lstset{basicstyle=\footnotesize}
\fbox{
\begin{lstlisting}[language=C++]
class Link_Field{
private:
	Afield links[SITES][NUMLINK];
public:
	Link_Field(void);
	Link_Field(int);
	Afield get(const Lattice_Vector &, const int) const;
	void set(const Lattice_Vector &, const int, const Afield &);
	void print(void);
	};
	
Link_Field Cjg(const Link_Field &);
Link_Field operator +(const Link_Field &, const Link_Field &);
Link_Field operator -(const Link_Field &, const Link_Field &);
Link_Field operator *(const double, const Link_Field &);
Link_Field operator *(const Complex &, const Link_Field &);
Complex operator *(const Link_Field &, const Link_Field &);
\end{lstlisting}}
\end{center}
\caption{The {\tt Link\_Field} class with overloading of operators (from {\tt utilities.h}).}
\label{fig:link-field}
\end{figure}

We look at the structure of the fermionic term $\eta \cDb_\mu \psi_\mu$ on the lattice and the structure of the corresponding fermionic operator in the code. On the lattice this fermionic term takes the form
\bea
\eta \cDb_\mu \psi_\mu &\rightarrow& \frac{1}{2} \Big[ \eta(\vx)\cDb_\mu^{(-)} \psi_\mu(\vx) + \psi_\mu(\vx) \cDb_\mu^{(+)} \eta(\vx)\Big]\nn \\
&=&\frac{1}{2} \Big[ \eta^a(\vx) T^a\cDb_\mu^{(-)}T^b \psi^b_\mu(\vx) + \psi^b_\mu(\vx) T^b\cDb_\mu^{(+)}T^a \eta^a(\vx)\Big]~,
\eea
where $T^a$ are the generators of the gauge group.

On expanding the lattice covariant difference operators we have
\bea
\eta \cDb_\mu \psi_\mu &\rightarrow&  \frac{1}{2} \Big[ \eta^a(\vx) T^a\Big(T^b \psi^b_\mu(\vx)\cU_\mu^{\dagger}(\vx) - \cU_\mu^{\dagger}(\vx - \hatbe_\mu)T^b \psi^b_\mu(\vx - \hatbe_\mu)\Big) \nn \\
&&+ \psi^b_\mu(\vx) T^b\Big( T^a \eta^a(\vx + \hatbe_\mu)\cU_\mu^{\dagger}(\vx) - \cU_\mu^{\dagger}(\vx)T^a \eta^a(\vx)\Big)\Big]\nn \\
&=& \frac{1}{2} \Big[ \eta^a(\vx) \Big(T^aT^b \cU_\mu^{\dagger}(\vx)\Big)\psi^b_\mu(\vx) - \eta^a(\vx)\Big(T^a\cU_\mu^{\dagger}(\vx - \hatbe_\mu)T^b \Big)\psi^b_\mu(\vx - \hatbe_\mu)\Big) \nn~~ \\
&&+ \psi^b_\mu(\vx) \Big(T^bT^a \cU_\mu^{\dagger}(\vx)\Big)\eta^a(\vx + \hatbe_\mu) - \psi^b_\mu(\vx)\Big(T^b\cU_\mu^{\dagger}(\vx)T^a\Big) \eta^a(\vx)\Big)\Big]~.~~~~~
\eea
In the code we compute the combination ${\rm Tr}(T^a\cU_\mu(\vx)T^b)$ as $V_\mu(\vx)^{ab}$ and store it as the object {\tt Adjoint\_Link\_Field}. It is this field that is passed into the functions that require the action of the twisted fermion operator in the inverter. Explicitly, the contribution to the operator coming from the term ${\rm Tr}~(\eta \cDb_\mu \psi_\mu)$ in the action takes the following form in the code:
\bec
\begin{tabular}{ l l}
{\tt +0.5*conjug(V.get(x,mu).get(a,b))} &  \\
{\tt -0.5*conjug(V.get(x-e\_mu,mu).get(b,a))*BC(x,-e\_mu)}&  \\
{\tt +0.5*conjug(V.get(x,mu).get(a,b))*BC(x,e\_mu)} & \\
{\tt -0.5*conjug(V.get(x,mu).get(b,a))} & \\
\end{tabular}
\eec

\section{Simulation results}
\label{sec:sim-results}

In this section we provide some numerical results obtained through the recent simulations of the two-dimensional $\cN=2$ lattice SYM theory \cite{Catterall:2011aa, Mehta:2011ud}.

The results we show in this section were obtained using the orbifold prescription for the parametrization of the complexified gauge fields $\cA_\mu(x)$ on the lattice. The continuum fields $\cA_a(x)$ are mapped to link fields $\cU_a(\vn)$ living on the link between $\vn$ and in $\vn+\hatbmu_a$ through the mapping:
\beq
\cU_a(\vn) = {\cA_a(\vn)}~,
\eeq
where $\cA_a(\vn)=\sum_{i=1}^{N_G} \cA_a^i T^i$ where $T^i=1 \ldots N_G$ are the anti-hermitian generators of a $U(N)$ group.  Notice though that in spite of the appearance of a complex connection the theory only possesses the usual $U(N)$ gauge symmetry. \footnote{Notice that our lattice gauge fields are dimensionless and hence contain an implicit factor of the lattice spacing $a$.} Simulations with linear gauge links of this type  have been investigated in \cite{Catterall:2011aa}.

\subsection{Eigenvalues of scalars}

The requirement that the lattice theory target the continuum theory as the lattice spacing is sent to zero demands vanishing of the fluctuations of all lattice fields and {\it in particular the fluctuations of the trace part of the scalar field $B^0_a$}. It is also important that the trace mode develops a nonzero expectation value of unity in order
that the lattice action yield the appropriate kinetic terms in the naive continuum limit.
Given the absence of any classical potential guaranteeing these features, we find that it is necessary to add a suitable gauge-invariant potential to the lattice theory to ensure these conditions hold\footnote{It was precisely this requirement that led to a truncation of the $U(N)$ symmetry to $SU(N)$ in the original simulations of these theories corresponding to a delta function potential for the $U(1)$ part of the field \cite{Catterall:2008dv}.}. In principle, once this mode is regulated one can examine whether this potential can be sent to zero in the continuum limit.

We have added a simple potential term of the following form to regulate the trace mode in the simulations\footnote{A potential term of this type was first introduced and tested in \cite{Hanada:2010qg}.}:
\beq
\label{eq:u1-mass-term}
S_M = \mu^2 \sum_\vx \left(\frac{1}{N}{\rm Tr}(\cU_a^\dagger(\vx) \cU_a(\vx))-1\right)^2~.
\eeq
This term fixes the vev of the scalar trace mode $B^0_a$  to unity and constrains the fluctuations of the trace mode $\delta B^0_a$ with a quadratic mass term at
leading order in the lattice spacing.
The remaining traceless fluctuations feel only a soft quartic potential.
\beq
S_M \approx \mu^2\sum_\vx [\delta B^0_a ]^2  + \ldots
\eeq
Since this $U(1)$ scalar sector decouples in the naive continuum limit this 
should not break the supersymmetry of the remaining $SU(N)$ sector for small enough lattice spacing (indeed all susy breaking terms should vanish as
$\mu\to 0$)

In the C++ code the mass term (\ref{eq:u1-mass-term}) is implemented using
\bec
{\tt (1.0/NCOLOR)*Tr(Udag.get(x,mu)*U.get(x,mu)).real()-1.0}
\eec
in {\tt action.cpp}. The $U(1)$ mass coefficient $\mu$ is denoted by the parameter {\tt BMASS} and should be held {\it fixed} as we take the continuum limit.
This implies that the physical mass is taken to infinity in this limit for any non-zero $\mu$ and hence that the expectation value of $B^0_a$ is frozen at unity in this
limit.

We rescale all lattice fields by powers of the lattice spacing to make them dimensionless. This leads to an overall dimensionless coupling parameter of the form $N/(2\lambda a^2)$, where $a=\beta/T$ is the lattice spacing, $\beta$ is the physical extent of the lattice in the Euclidean time direction and $T$ is the number of lattice sites in the time-direction. The coupling $\lambda = g^2N$ is the usual 't Hooft parameter. Thus, the lattice coupling
\beq
\label{eq:lattice-coupling}
\kappa=\frac{NT^2}{2\lambda\beta^2}~,
\eeq
for the symmetric two-dimensional lattice where the spatial length $L=T$\footnote{To obtain this dimensionally reduced model from the $\cN=4$ theory
one merely sets the parameters ${\tt LX}=1$, ${\tt LY}=1$ in {\tt utilities.h}.} Note that $\lambda\beta^2$ is the dimensionless physical `t Hooft coupling measured in units of the area. In these two dimensional simulations, the continuum limit can be approached by fixing $t=\lambda\beta^{2}$ and $N$, and increasing the number of lattice points $L \rightarrow \infty$. We have taken three different values for this coupling  $t=0.5, 1.0, 2.0$ and lattice sizes ranging from $L=2, \cdots, 12$. In Fig. 6 we show the average scalar eigenvalue given by $\cU_a^\dagger \cU_a - I$ for the $\cQ=4$ model as a function of the lattice size $L$. This figure confirms that as $L \rightarrow \infty$ we are indeed approaching a continuum limit since the scalar eigenvalues (which contain a factor of $a$ to render them dimensionless) are driven to zero. 
\begin{figure}
\centering
\includegraphics[width=10cm]{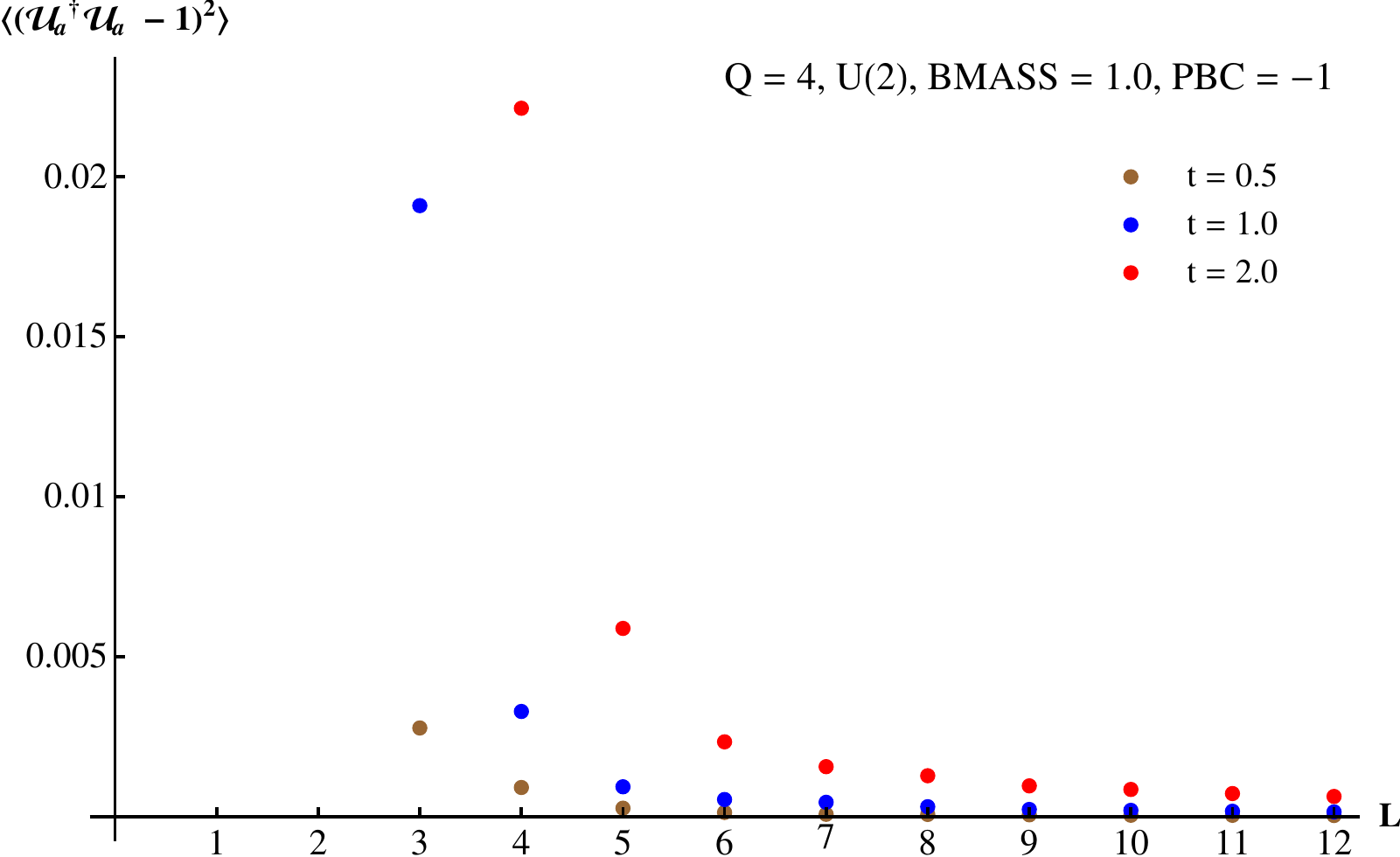}
\label{fig:q4u2scalar}
\caption{Plot showing the average scalar eigenvalue versus the lattice size $L$ in the two-dimensional $\cQ=4$ theory.}
\end{figure}

\subsection{Pfaffian phase/sign problems}

The models we have discussed may encounter an additional difficulty in the context of simulation - the fermionic sign problem. After integration over the fermions the effective bosonic action picks up a contribution from the logarithm of the fermionic Pfaffian ${\rm Pf}(M)$ which is not necessarily real. Indeed for the supersymmetric lattice constructions we described above, $M$ at non zero lattice spacing is a complex operator and one might worry that the resulting Pfaffian could exhibit a fluctuating phase $e^{i\alpha}$. Since Monte Carlo simulations must necessarily be performed with a positive definite measure the only way to incorporate this phase is through a re-weighting procedure which folds the phase in with the observables of the theory. Expectation values of observables derived from such simulations can then suffer huge statistical errors which swamp the signal rendering the Monte Carlo techniques effectively useless.

In Fig. 7 we show results for $\langle|\sin(\alpha)|\rangle$ as a function of $L$ for the $\cQ=4$ model with gauge group $U(2)$ (edit {\tt utilities.h} to change number
of supercharges). Three values of $t=\lambda\beta^2$ are shown in each plot but the behavior is qualitatively similar for all $t$. We have used the mass parameter controlling the $U(1)$ mode as {\tt BMASS} = 1. These numerical results show that while this model appears to suffer from a  sign problem for coarse lattices these effects disappear as the lattice is refined and the phase fluctuations are driven to zero as the continuum limit is taken. This is consistent with the work reported  in \cite{Hanada:2010qg}\begin{figure}
\centering\includegraphics[width=10cm]{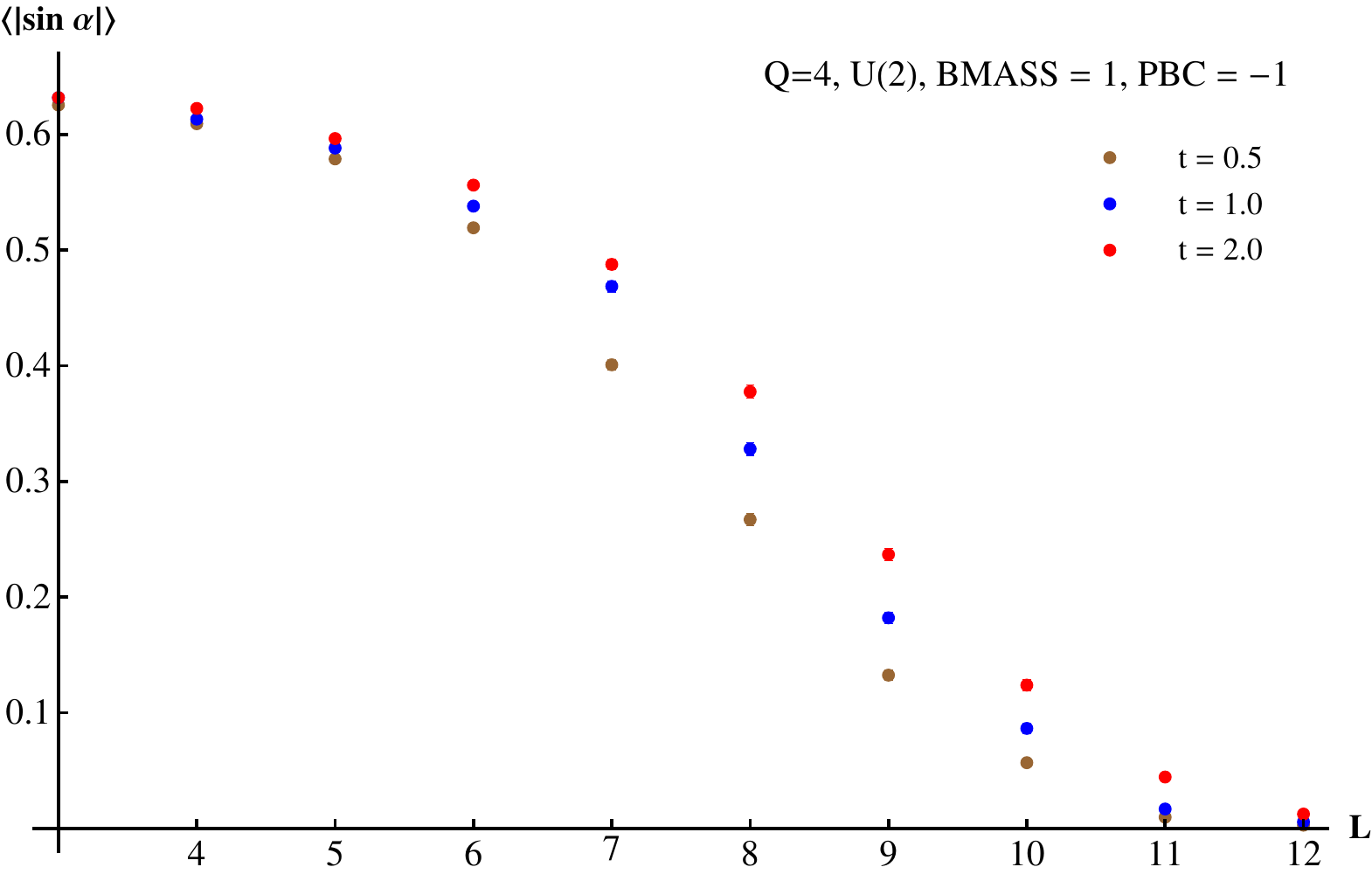}
\caption{Plot showing the average of the $\sin$ of the Pfaffian phase $\alpha$ against the lattice size $L$ in the $\cQ=4$ lattice SYM with gauge group $U(2)$ and exponential representation for the gauge links.}
\end{figure}

\subsection{Restoration of supersymmetry}

The topological nature of the twisted theory formulated on a torus with periodic boundary conditions can be used to show that the partition function of the lattice model is actually independent of the coupling constant. Thus derivatives of the partition function with respect to the coupling constant such as the expectation value of the action must vanish. Since the fermions enter only quadratically, their contribution can be evaluated simply using a scaling argument and thence a simple expression derived for the expectation value of the bosonic action. Thus measurements of $\langle S_B(\cU , \cUb)\rangle$ provide us with a check that the scalar supersymmetry has indeed been implemented correctly in our codes. Actually, since in practice we use supersymmetry breaking (thermal) boundary conditions (and also employ a supersymmetry breaking potential for the scalar $U(1)$ mode) to do simulations, measuring this quantity provides some insight into the magnitude of supersymmetry breaking effects in the theory.  

In the case of two-dimensional $\cQ=4$ theory, we have the expression for the mean action 
\beq
\langle S \rangle = -\frac{\partial \ln Z}{\partial \kappa} = \langle \cQ \Lambda \rangle = 0~,
\eeq
where $\kappa$ is the coupling constant of the twisted action and the last equality follows from the $\cQ$-exact nature of the twisted theory and shows that the vanishing mean action can be thought of as arising as a consequence of a simple $\cQ$-Ward identity.

If we integrate out the twisted fermions and the auxiliary field $d$ we find the following expression for the partition function of the two-dimensional $\cQ=4$ theory
\beq
Z = \kappa^{4N_G V/2} \kappa^{-N_GV/2} \int D\cU D\cUb e^{-\kappa S_B(\cU, \cUb)}{ \det}^{\frac{1}{4}} (M(\cU, \cUb))~,
\eeq
where $N_G$ is the number of generators of the gauge group and $V$ is the number of lattice points. The first pre-factor arises from the fermion integration and the second derives from the Gaussian integration over the auxiliary field. From this we find the following condition on the mean bosonic action as a consequence of the scalar supersymmetry $\cQ$:
\beq
\label{eq:exact-baction}
\langle \kappa S_B \rangle = \frac{3}{2}N_G V~.
\eeq
In Fig. 8 we show the mean bosonic action on the lattice against the lattice size $L$. The thick solid line represents the exact value of the bosonic action given in \ref{eq:exact-baction}.
\begin{figure}
\centering\includegraphics[width=10cm]{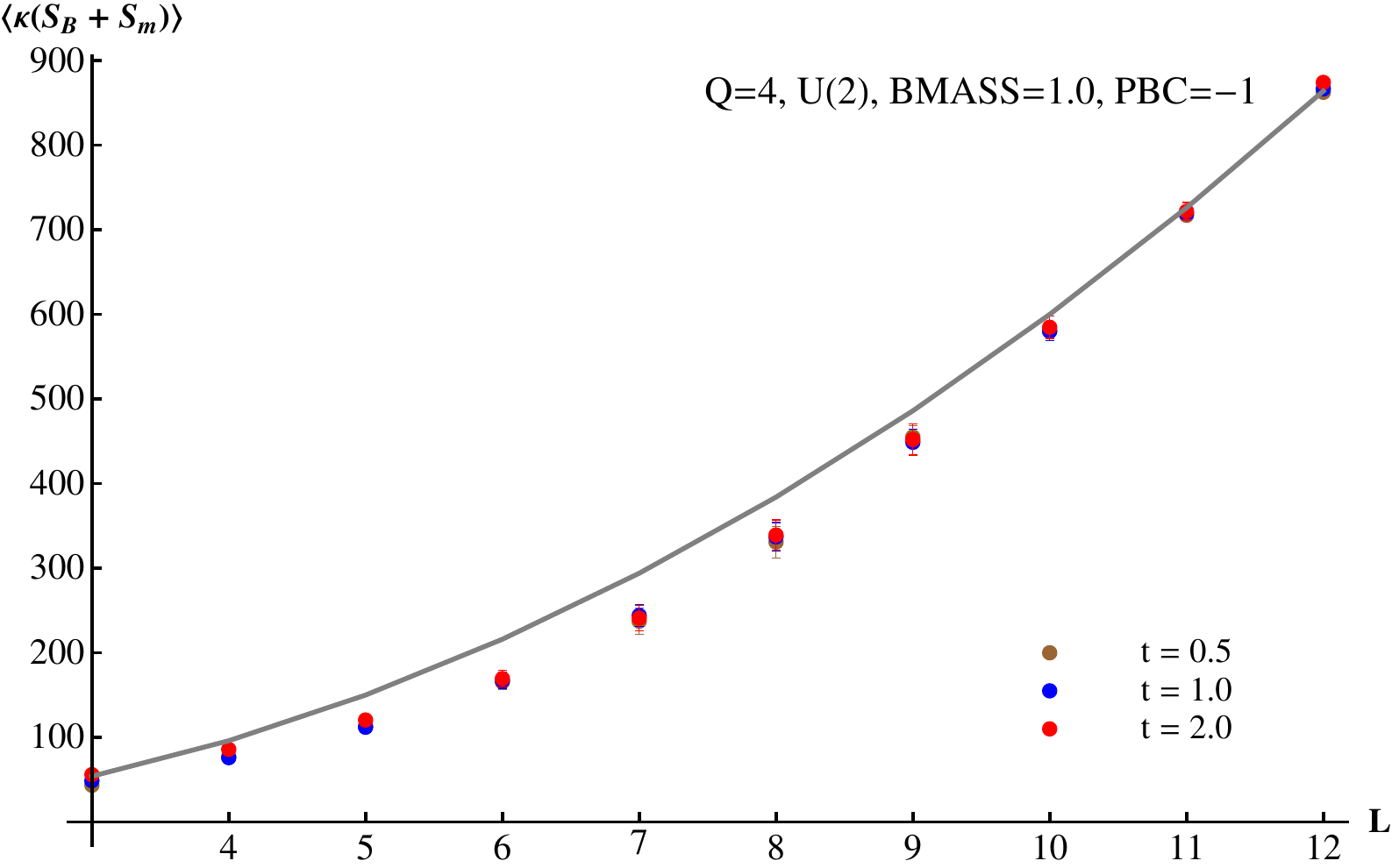}
\caption{Plot showing the average bosonic action $\langle \kappa (S_B+S_M) \rangle$ on the lattice against the lattice size $L$ in the $\cQ=4$ lattice SYM with gauge group $U(2)$ and exponential parametrization for the gauge links. The thick solid line corresponds to the exact value of the bosonic action.}
\end{figure}
Clearly the lattice measurements approach the exact result for sufficiently small lattice spacing. The deviations that are visible are presumably related to the fact that we have a sign problem (these measurements do {\it not} incorporate re-weighting) for small $L$ and the simulations are also conducted at non zero temperature. We have shown that the sign problem disappears in the continuum limit which is consistent with the much better agreement at large $L$. To recover the true zero temperature result requires in principle that we extrapolate our measurements to $t\to\infty$ after taking the thermodynamic limit.
\section{Conclusions and outlook}
\label{sec:conclu}
In this paper we have described in some detail the construction of an object oriented code suitable for the simulation of a recently discovered class of lattice field theories possessing exact supersymmetry. The continuum construction and lattice discretization of $\cQ=4,8,16$ supercharge SYM theories in two, three and four dimensions are all covered in detail. The structure of the problem requires the construction of unusual data structures for representing the fermions, which is the primary difference between the code described here and more conventional codes suitable for simulating QCD. Nevertheless the basic algorithms employed (RHMC and multi-mass CG solvers) are borrowed directly from lattice QCD and adapted to the problem at hand. We verify the correctness of the resultant code by showing results from simulations of the two-dimensional SYM model. Acceleration of this code can be achieved by off-loading the linear solver calculation to a GPU card - we refer the reader to \cite{Galvez:2011cd} for details. It is also possible to parallelize the code with suitable distributed libraries layered over MPI \cite{DiPierro:2005jd} and work in both these directions is ongoing.

\section{Acknowledgments}
This work is supported in part by DOE under grant number DE-FG02-85ER40237. Simulations were performed using USQCD resources at Fermilab. We would like to thank useful discussions with Richard Galvez, Joel Giedt, Dhagash Mehta and Greg van Anders.

\appendix

\section{Installation of the program}
\label{sec:install}
It is very easy to perform the installation and execution of SUSY\_LATTICE. Below we provide the necessary steps on Unix or Linux systems.

\begin{itemize}
\item[1.] Download the code from CPC Program Library and unpack it.
\item[2.] Change the directory to SUSY\_LATTICE.
\item[3.] Compile the code (g++ -O *.cpp -o SUSY\_LATTICE -llapack -lblas).
\item[4.] Modify the input parameters located in file {\it parameters}
\item[5.] Type ./SUSY\_LATTICE $>$\& log \& to run the code.
\end{itemize} 

The authors have tested the code on Linux machines. After slight modifications of above steps the code may be installed on other machines.

The output of the code produces the following files in the running directory: 
\begin{itemize}
\item[1.] {\tt cgs:} Average number of conjugate gradient (CG) iterations. (See {\tt MCG\_solver.cpp}).
\item[2.] {\tt config:} File to read in containing the site, link and plaquette field configurations from a previous run. (See {\tt read\_in.cpp}.)
\item[3.] {\tt corrlines:} Correlation function between temporal Polyakov lines as function of spatial separation (See {\tt corrlines.cpp}.)
\item[4.] {\tt data:} Boson (1st column) and fermion (2nd column)  contributions to the total action. (See {\tt measure.cpp}.)
\item[5.] {\tt dump:} Site, link and plaquette field configurations stored as ASCII (See {\tt write\_out.cpp}.)
\item[6.] {\tt eigenvalues:}  Eigenvalues of $\cU^dagger_a(x)\cU_a(x)$ $N$ real numbers for each lattice point $x$ and direction $a$ (See {\tt measure.cpp}.)
\item[7.] {\tt hmc\_test:} $e^{-Delta H}$ from HMC test (See {\tt update.cpp}.)
\item[8.] {\tt lines\_s:} Spatial Polyakov line. (See {\tt measure.cpp}.)
\item[9.] {\tt lines\_t:} Temporal Polyakov line. (See {\tt measure.cpp}.)
\item[10.] {\tt log:} Log file.
\item[11.] {\tt loops:} Wilson loops. (See {\tt loop.cpp}.)
\item[12.] {\tt scalars:} ${\rm Tr} (U^\dagger U)$ (See {\tt measure.cpp}.)
\item[13.] {\tt ulines\_s:} The spatial Polyakov line computed using the unitary part of the link (See {\tt measure.cpp}.)
\item[14.] {\tt ulines\_t:} The temporal Polyakov line computed using the unitary part of the link (See {\tt measure.cpp}.)
\end{itemize}

\section{The list of files in SUSY\_LATTICE library}
\label{sec:file-list}
We list the files included in SUSY\_LATTICE library with a brief description of their purpose.
\bec
\begin{itemize}
\item[1.]{\tt action.cpp:} Compute the total action - fermionic and bosonic.
\item[2.]{\tt corrlines.cpp:} Finds the traced product of the link matrices at various lattice sites.
\item[3.]{\tt evolve\_fields.cpp:} Leapfrog evolution algorithm. Also stores the fermion and boson forces for the next iteration.
\item[4.]{\tt fermion\_forces.cpp:} Computes the fermion kick to gauge link force.
\item[5.]{\tt force.cpp:} Bosonic and pseudo-fermionic contribution to the force.
\item[6.]{\tt kinetic\_energy.cpp:} Computes the kinetic energy term in the Hamiltonian.
\item[7.]{\tt line.cpp:} Computes the Polyakov lines.
\item[8.]{\tt loop.cpp:} Computes the Wilson loops.
\item[9.]{\tt matrix.cpp:} Builds the fermion matrix (sparse and full forms) and also computes the Pfaffian of the fermion operator.
\item[10.]{\tt MCG\_solver.cpp:} multi-mass CG solver needed for RHMC alg.
\item[11.]{\tt measure.cpp:} Performs measurements on field configurations. Writes out scalar eigenvalues, Polyakov/Wilson loops and the action.
\item[12.]{\tt my\_gen.cpp:} Computes $SU(N)$ generator matrices.
\item[13.]{\tt obs.cpp:} Computes fermion and gauge actions. Also returns the unitary piece of the complex link field.
\item[14.]{\tt read\_in.cpp:} Reads in the previously generated field configurations - file {\tt config}
\item[15.]{\tt read\_param.cpp:} Reads in the simulation parameters from a data file called {\tt parameters}
\item[16.]{\tt setup.cpp:} Contains the partial fraction coefficients necessary to represent fractional power of fermion operator - used  by Remez algorithm.
\item[17.]{\tt sym.cpp:} The main program - performs warm up on field configurations and commences measurement sweeps once the configurations are warmed up.
\item[18.]{\tt unit.cpp:} Extracts the unitary piece of the complex gauge links.
\item[19.]{\tt update.cpp:} Updates the field configurations based on HMC test.
\item[20.]{\tt utilities.cpp:} Utility functions. Contains constructors for site, link, plaquette fields, gauge fields, twist fermions etc. Edit to change
number of supercharges and size of lattice dimensions.
\item[21.]{\tt write\_out.cpp:} Writes out the values of gauge and twist fermion fields on to a file called {\tt dump}.
\end{itemize}
\eec

\section{A sample input parameter file for SUSY\_LATTICE}
\label{sec:input-param}

This is a sample input parameter file called {\tt parameters} located in the SUSY\_LATTICE folder.
\bec
\begin{tabular}{ l l l l l l l l }
${\tt 10000}$ &  ${\tt 50}$  &  ${\tt 10}$ &  ${\tt 0.5}$  &   ${\tt 1.0}$ &  ${\tt 0.02}$ &  ${\tt 0.0}$  &  ${\tt 0}$ \\
 & & & & & & & \\
{\tt SWEEPS} & {\tt THERM} & {\tt GAP} & {\tt LAMBDA} & {\tt BETA} & {\tt DT} &  {\tt ALPHA} & {\tt READIN} \\
\end{tabular}
\eec
There are the definitions of the parameters:
\begin{itemize}
\item[1] {\tt SWEEPS:} Total number of Monte Carlo time steps intended for taking measurement steps.
\item[2.] {\tt THERM:} Total number of Monte Carlo time steps intended for thermalizing the field configurations.
\item[3.] {\tt GAP:} The gap between measurement steps.
\item[4.] {\tt LAMBDA:} The `t Hooft coupling.
\item[5.] {\tt BETA:} Inverse temperature.
\item[6.] {\tt DT:} The time step put in the integrator for leapfrog evolution.
\item[7.] {\tt ALPHA:} A supersymmetric mass (deformation) parameter.
\item[8.] {\tt READIN:} Determines whether to read in the previously generated field configurations or not. The program will read in the previous configurations if {\tt READIN} is set to ${\tt 1}$. 
\end{itemize}

\newpage

\noindent{\bf PROGRAM SUMMARY}\\
\begin{small}
{\em Manuscript Title:}~ An object oriented code for simulating supersymmetric Yang--Mills theories \\
{\em Authors:}~ Simon Catterall and Anosh Joseph \\
{\em Program Title:}~ SUSY\_LATTICE \\
{\em Journal Reference:}                                      \\
%
{\em Catalogue identifier:}                                   \\
%
{\em Licensing provisions:}~ None\\
%
{\em Programming language:}~C++\\
{\em Operating system:}~Any, tested on Linux machines\\
%
{\em Keywords:}~Lattice Gauge Theory \sep Supersymmetric Yang--Mills \sep Rational Hybrid Monte Carlo \sep Object Oriented Programming  \\
{\em PACS:}~11.15.Ha, 12.60.Jv, 12.10.-g, 12.15.-y, 87.55.kd, 87.55.kh\\
{\em Classification:}~11.6 Phenomenological and Empirical Models and Theories\\
{\em Nature of problem:}\\
To compute some of the observables of supersymmetric Yang--Mills theories such as supersymmetric action, Polyakov/Wilson loops, scalar eigenvalues and Pfaffian phases.  \\
{\em Solution method:}\\ 
We use the Rational Hybrid Monte Carlo algorithm followed by a Leapfrog evolution and a Metroplois test. The input parameters of the model are read in from a parameter file.\\
{\em Restrictions:}\\
This code applies only to supersymmetric gauge theories with extended supersymmetry, which undergo the process of maximal twisting. (See Section \ref{sec:method-twist-SYM} of the manuscript for details.)\\
{\em Unusual features:}\\
{\em Running time:}\\
From a few minutes to several hours depending on the amount of statistics needed.\\
{\em References:}

\end{small}

\end{document}